\documentclass[aps,prl,amssymb,amsmath,superscriptaddress,twocolumn,floatfix,showpacs]{revtex4} 
\pdfoutput=1

\usepackage{amsmath}
\usepackage{amsfonts}
\usepackage{amssymb}
\usepackage{mathbbol}
\usepackage{natbib}
\usepackage{color}
\usepackage[bookmarks, pdftitle={Strong No-Go Theorem for Gaussian Quantum Bit Commitment}, pdfauthor={Loick Magnin, Frederic Magniez, Anthony Leverrier, Nicolas J. Cerf}]{hyperref}

\def\Tr{{\rm Tr}}
\def\cqfd{\hfill$\square$}
\def\FF{{\cal F}}

\def\HH{{\cal H}}
\def\LL{{\cal L}}
\def\EE{{\cal E}}
\def\SS{\mathcal{S}}

\def\I{\mathbb{I}}
\def\R{\mathbb{R}}

\newcommand{\ket}[1]{| #1 \rangle}
\newcommand{\bra}[1]{\langle #1 |}
\newcommand{\braket}[2]{\langle #1 | #2 \rangle}
\newcommand{\proj}[1]{| #1 \rangle\!\langle #1 |}
\newcommand{\abs}[1]{\left| #1 \right|}
\newcommand{\norm}[1]{\left\| #1 \right\|}
\newcommand{\inv}[1]{\frac{1}{ #1 }}

\newcommand{\comment}[1]{\textsf{[#1]}}
\newcommand{\TODO}[1]{\textsf{[TODO: #1]}}
\renewcommand{\comment}[1]{}
\renewcommand{\TODO}[1]{}

\newtheorem{thm}{Theorem}
\newtheorem{lem}{Lemma}

\begin{document}
\title{Strong No-Go Theorem for Gaussian Quantum Bit Commitment}

\author{Lo\"ick Magnin}
\affiliation{QuIC, \'Ecole Polytechnique, CP 165,
Universit\'e Libre de Bruxelles, 1050 Brussels, Belgium}
\affiliation{LRI, Univ Paris-Sud, CNRS; F-91405 Orsay, France}

\author{Fr\'ed\'eric Magniez}
\affiliation{LRI, Univ Paris-Sud, CNRS; F-91405 Orsay, France} 

\author{Anthony Leverrier}
\affiliation{Institut Telecom /
  Telecom ParisTech, CNRS LTCI, 46 rue Barrault, 75634 Paris
  Cedex 13, France}

\author{Nicolas J. Cerf}
\affiliation{QuIC, \'Ecole Polytechnique, CP 165, 
Universit\'e Libre de Bruxelles, 1050 Brussels, Belgium}
\affiliation{Research Laboratory of Electronics, Massachusetts Institute of Technology,
Cambridge, Massachusetts 02139, USA}

\date{\today}

\begin{abstract}
Unconditionally secure bit commitment is forbidden by quantum mechanics. We extend this no-go theorem to continuous-variable protocols where both players are restricted to use Gaussian states and operations, which is a reasonable assumption in current-state optical implementations.  Our Gaussian no-go theorem also provides a natural counter-example to a conjecture that quantum mechanics can be rederived from the assumption that
key distribution is allowed while bit commitment is forbidden in Nature.
\end{abstract}



\maketitle

\phantomsection
\addcontentsline{toc}{section}{Introduction}

Bit commitment is a cryptographic primitive with a large scope of applications ranging from two-party secure computation, e.g., secure authentication, to coin flipping. It involves two mistrustful parties: Alice must commit to a certain bit, which should remain hidden to Bob until she reveals its value. A traditional picture for this protocol is as follows: Alice locks a secret bit into a safe that she gives to Bob; then, when she wants to reveal her secret, she simply hands over the key of the safe to Bob. A bit commitment protocol is said to be secure if it prevents Alice to cheat (i.e., she cannot change the value of the bit she had commited) and Bob to cheat (i.e., he cannot learn information about the bit before Alice reveals it).

This primitive has been exhaustively studied in classical cryptography, where the security relies on unproven computational assumptions \cite{Nao91, Cha86}. The idea of quantum bit commitment (QBC) was first introduced by Bennett and Brassard in 1984 \cite{BB84}, together with the famous BB84 quantum key distribution protocol. In 1993, Brassard \textit{et al.} proposed a QBC protocol known as BCJL \cite{BCJL93}, which was believed to be secure until 1996, when Mayers \cite{May97} and independently Lo and Chau \cite{LC97} proved that it was not the case. Their proof involved a reduction of the BCJL protocol to a purified protocol, which cannot be perfectly secure against both Alice and Bob. Thus, it appeared that this reduction precludes the existence of an unconditionally secure QBC protocol. Because of the complexity of this reduction, however, it was not universally accepted (see, e.g., \cite{Yue00}) until 2006, when d'Ariano \textit{et al.} provided a complete, formal description of QBC protocols that definitely closed the question \cite{AKSW07}. This is the content of the \textit{no-go theorem} for QBC.

Interestingly, this situation is in sharp contrast with quantum key distribution, for which unconditionally secure protocols have been exhibited \cite{SBCDLP08}. These two facts, namely the possibility of key distribution and impossibility of bit commitment, seem to be specific to quantum mechanics, and were conjectured by Brassard and Fuchs to be actually sufficient to rederive quantum mechanics from first principles \cite{Bra05}. This conjecture was later proven wrong, but Clifton \textit{et al.} proved instead that the assumptions of no-signalling, no-broadcasting, and the impossibility of bit commitment make it work within the framework of $C^*$-algebras \cite{CBH03}. This
is known as the CBH theorem.

Coming back to the no-go theorem for QBC, let us stress that it only applies to unconditionally secure protocols, that is, to the case where Alice and Bob have no restriction on their capabilities except those dictated by quantum mechanics. This leaves the door open to QBC protocols that could be secure under reasonable assumptions on Alice and Bob's capabilities. Such protocols were found in the bounded-storage model \cite{DFSS08} or by exploiting the constraints imposed by special relativity \cite{Ken99}.


In this Letter, we address QBC protocols with \textit{continuous variables}, and explore whether such protocols may be found secure when both parties are restricted to use Gaussian states and operations. Most quantum information protocols to date have been based on discrete variables in a finite-dimensional Hilbert space. Recently, however, continuous variables (CV) have been proven to be a very powerful alternative approach \cite{CLP07}. In the case of optical communication, for example, the quadrature components of the light field make especially useful continuous variables because of their associated detection scheme, namely homodyne detection.
This is well illustrated with CV quantum key distribution, which was recently proven unconditionaly secure \cite{RC09} and appears as a credible alternative to single-photon based quantum key distribution \cite{LG09}.
Dealing with CV quantum information protocols unfortunately comes with a price, namely that their analysis may be intractable as an infinite-dimensional Hilbert space is involved. 

An elegant solution consists in restricting the analysis to so-called \textit{Gaussian states and operations}, which, apart from being efficiently characterizable within the appropriate formalism, can be relatively easily manipulated in the laboratory. It is therefore a very natural and important question to ask whether QBC protocols can be built with continuous variables, which could be made secure 
if both parties are capable to manipulate Gaussian states only.
Remember that, although the no-go theorem for unconditional security holds for infinite-dimensional Hilbert spaces as such, it is unknown
whether secure QBC can exist when both parties have restricted capabilities. Here, we answer by the negative if this restriction is put at the boundary of the set of Gaussian states, and establish a \textit{strong} no-go theorem for Gaussian QBC protocols. Specifically, for any Gaussian QBC protocol, we find a corresponding Gaussian cheating strategy. 
Moreover, we provide a constructive attack for any CV QBC protocol, whereas constructive attacks were previously known for finite dimensions only.


Let us first recall some notions linked to the distinguishability of quantum states. 
The \emph{fidelity} between two states $\rho$ and $\sigma$ is defined as $\FF(\rho,\sigma)=\left(\Tr\sqrt{\sqrt{\rho}\sigma\sqrt{\rho}}\right)^2$.
If $\rho = \proj{\psi}$ and $\sigma = \proj{\phi}$ are pure states, the fidelity is simply $\abs{\braket{\psi}{\phi}}^2$.
Any purifications $\ket{\psi}$ of $\rho$ and $\ket{\phi}$ of $\sigma$ satisfy $\FF(\psi,\phi) \leq \FF(\rho,\sigma)$.
Uhlmann's theorem \cite{Uhl76} states that this inequality can always be saturated, that is, there exists a purification of $\rho$ (resp. $\sigma$) noted $\ket{\psi'}$ (resp. $\ket{\phi'}$) which is such that $\FF(\psi',\phi')=\FF(\rho,\sigma)$. Although this has been shown regardless of the dimension, constructive proofs of this purification are known in finite dimensions only \cite{Joz94}. 
The \emph{trace distance} between the states $\rho$ and $\sigma$ is defined as $D(\rho,\sigma) = \inv{2}\norm{\rho-\sigma}_{1}$, where $\norm{\tau}_{1} = \Tr\sqrt{\tau^\dagger\tau}$ for any operator $\tau$. The trace distance is related to the \textit{guessing probability} $\inv{2}(1+D(\rho,\sigma))$, which is the maximum probability of distinguishing the two states with the best measurement.
We also recall a useful relation between the fidelity and trace distance,
\begin{align}
	D(\rho,\sigma) &\leq \sqrt{1-\FF(\rho,\sigma)} ,
	\label{fidelity-distance}
\end{align}
as well as the Bhattacharyya bound \cite{Kai67,PL08}, namely
\begin{align}
  1-D(\rho,\sigma)\leq\Tr(\sqrt{\rho}\sqrt{\sigma}) .
	\label{battacharyya}
\end{align}


\phantomsection
\addcontentsline{toc}{section}{Quantum Bit Commitment}

\emph{Quantum bit commitment} ---  
Formally, any (reduced) QBC protocol can be described as follows: Alice encodes her bit $b$ into a pure bipartite state $\ket{\psi_b}$ ans sends one half to Bob. At the end of the committing phase, Bob holds either $\rho_0 = \Tr_{A}\proj{\psi_0}$ or $\rho_1 = \Tr_{A}\proj{\psi_1}$ if Alice wants to commit to $0$ or $1$, respectively. The protocol is referred to as \emph{$\varepsilon$-concealing} if $D(\rho_{0},\rho_{1})\leq\varepsilon$, which means that Bob cannot learn the value of $b$, except with probability $\varepsilon$. To reveal her bit, Alice sends the other half of $\ket{\psi_{b}}$. In a so-called \emph{$\delta$-cheating strategy}, Alice sends a state $\rho^\sharp$ in the committing phase and then decides to follow a strategy leading to a final state of her choice, $\ket{\psi^\sharp_0}$ or $\ket{\psi^\sharp_1}$, so that Bob should not be able to distinguish this strategy from a honest strategy with a probability greater than $\delta$. This means that $D(\rho^\sharp,\rho_b)\leq\delta$ and $D(\psi^\sharp_b,\psi_b)\leq\delta$.
Here, we will only consider the simple strategy in which $\rho^\sharp = \rho_0$ and $\ket{\psi^\sharp_0} = \ket{\psi_0}$. Thus, $\ket{\psi^\sharp_1}$ 
will correspond to Alice initially committing to a zero and then cheating so to make it a one.
Without loss of generality, we will also consider that $\ket{\psi_b}$ are $2n$-mode states and $\rho_b$ are $n$-mode states.
Now, let us state our main result:
\begin{thm}
	\label{thethm}
	Given any $\varepsilon$-concealing Gaussian quantum bit commitment protocol to Bob, there exists a Gaussian $\sqrt{2\varepsilon}$-cheating strategy for Alice.
\end{thm}

For finite-dimensional protocols, the cheating strategy is usually exhibited with the help of Uhlmann's theorem, which gives purifications $\ket{\psi_{0}}$ of $\rho_{0}$ and $\ket{\psi_{1}}$ of $\rho_{1}$ such that $\FF(\psi_0,\psi_1)=\FF(\rho_0,\rho_1)$. 
Unfortunately, it is not known how to use this theorem to explicitly construct such purifications in infinite dimensions, and, even so, it would not help making statements about the Gaussianity of such purifications for Gaussian states. Our approach is based instead on the notion of \emph{intrinsic purification}, for which we give an explicit construction guaranteeing that every Gaussian state has a Gaussian intrinsic purification. Although this purification does not reach Uhlmann's bound, we derive an inequality which is sufficient to prove our theorem:



\begin{lem}
	\label{lemkey}
	Given the $n$-mode states $\rho_{0}$ and $\rho_{1}$, there exist $2n$-mode purifications $\ket{\hat{\psi}_{0}}$ of $\rho_{0}$ and $\ket{\hat{\psi}_{1}}$ of $\rho_{1}$ such that
	\begin{align}
		D(\hat{\psi}_{0},\hat{\psi}_{1}) \leq \sqrt{2 \, D(\rho_{0},\rho_{1})}.
	\end{align}
	Moreover, if $\rho_{0}$ and $\rho_{1}$ are Gaussian states, so are their purifications $\ket{\hat{\psi}_{0}}$ and $\ket{\hat{\psi}_{1}}$.
\end{lem}


\phantomsection
\addcontentsline{toc}{section}{Gaussian Formalism}

\emph{Gaussian formalism} --- 
The state $\rho$ of an $n$-mode bosonic quantum system is a unit-trace Hermitian positive semi-definite operator 
on $\HH^{\otimes n}$, where $\HH$ is the infinite-dimensional Hilbert space spanned by the excitations of each mode.
We note $\mathbf{i}=i_1\dots i_n$ and $\ket{\mathbf{i}}=\ket{i_1}\cdots\ket{i_n}$,
where $\{\ket{i}\}$ is the Fock basis of $\HH$. Since $\HH$ is isomorphic to $\LL^2(\R)$, any state $\rho$ is completely characterized by its Wigner function $W_{\rho}$, a quasi-probability distribution in the $2n-$dimensional phase space parametrized by the vector of quadratures $\xi = (x_1,p_1,\ldots,x_n,p_n)$. The covariance matrix $\gamma$ of $W_{\rho}$ is a  real, symmetric and positive matrix satisfying the Heisenberg inequality  $\gamma + i\Omega \geq 0$ where $\Omega = \bigoplus_{k=1}^n  \left[\begin{smallmatrix}0&1\\-1&0 \end{smallmatrix} \right]$.
An $n$-mode state is called \emph{Gaussian} if its Wigner function is Gaussian,
\begin{align}
W_{\rho}(\xi) = \inv{(2\pi)^n \sqrt{\det \gamma}}\exp\left\{ -\inv{2}(\xi-\mu)^T\gamma^{-1}(\xi-\mu)\right\}.\nonumber
\end{align}
Note that a Gaussian state is fully described by its first- and second-order moments $\mu\in\R^{2n}$ and $\gamma\in\R^{2n}\times\R^{2n}$.

A \emph{Gaussian operation} $\EE$ maps any Gaussian state to a Gaussian state. Therefore, $\EE$ is fully characterized by its action on the first- and second-order moments of a state. Furthermore, $\EE$ is a Gaussian unitary operator if and only if there exists a symplectic matrix $S$ (such that $S\Omega S^T = \Omega$) and a displacement vector $d$ such that for all states $\rho$, $W_{\EE(\rho)}(\xi) = W_{\rho}(S^{-1}\xi - d)$ \cite{ADMS95}.
The Williamson decomposition theorem states that a covariance matrix $\gamma$ is described by its \emph{symplectic eigenvalues} $\{\nu_{1},\dots,\nu_{n}\}$. More specifically, for any $\gamma$, there exists a symplectic transformation $S$ such that $S\gamma S^T = \bigoplus_{k=1}^n \nu_{k} \I_2$, with $\nu_{k}\geq 1$ \cite{SCS99}. In particular, for a Gaussian state $\rho$, there exists a Gaussian operation $V$, a \emph{Williamson unitary}, such that $V^{\dagger} \rho V =  \sum_{\mathbf{i}} \left( \prod_{k=1}^n (1-x_{k})x_{k}^{i_{k}} \right) \proj{\mathbf{i}}$, where $x_k = \frac{\nu_{k}-1}{\nu_{k}+1}$. In other words, any Gaussian state $\rho$ can be mapped via a Gaussian operation $V$ onto a tensor product of thermal states with symplectic eigenvalues $\nu_k$.

\phantomsection
\addcontentsline{toc}{section}{Gaussian Intrinsic Purification}

\emph{Gaussian intrinsic purification} --- 
Let $\rho$ be a $n$-mode state and $U$ be a diagonalization of $\rho$ in the Fock basis, that is, $U$ is a unitary operator such that $\bra{\mathbf{i}} U^\dagger\rho U\ket{\mathbf{j}} = p_{\mathbf{i}} \, \delta_{\mathbf{i}\mathbf{j}}$, where $\delta_{\mathbf{i}\mathbf{j}}$ is the Kronecker delta.
We then define an intrinsic purification $\ket{\hat{\psi}}$ of $\rho$ as
\begin{align}
	\ket{\hat{\psi}} = (U^* \otimes U) \sum_{\mathbf{i}} \sqrt{p_{\mathbf{i}}}\ket{\mathbf{i}}\ket{\mathbf{i}}.
\end{align}
(Note that it is not unique.)
Here and in what follows, $A^{*}$ (resp. $A^T$) denotes the complex conjugate (resp. the transpose) of any linear operator $A$ relatively to the Fock basis,
defined as $\bra{\mathbf{i}}A^*\ket{\mathbf{j}} =  \bra{\mathbf{i}}A\ket{\mathbf{j}}^*$ and $\bra{\mathbf{i}}A^T\ket{\mathbf{j}} = \bra{\mathbf{j}}A\ket{\mathbf{i}}$. 

A \emph{Gaussian intrinsic purification} of a Gaussian state $\rho$ thus consists in choosing $U = V$, that is, using a Williamson unitary in order to diagonalize $\rho$ in the Fock basis. Let us show that this purification is indeed Gaussian. The state $\sum_{\mathbf{i}} \sqrt{p_{\mathbf{i}}}\ket{\mathbf{i}}\ket{\mathbf{i}}$, being a tensor product of two-mode squeezed states, is Gaussian. Since $U$ is a Gaussian unitary operator,
all is left to show in order to prove that $\ket{\hat{\psi}}$ is a Gaussian state
is that $U^*$ is a Gaussian unitary operator too.
Let us take an arbitrary $n$-mode Gaussian state $\tau$ and assume that $U$ is described by the symplectic matrix $S$ and displacement vector $d$. 
We want to show that applying $U^*$ to $\tau$ is equivalent 
to applying the symplectic matrix $\Sigma_{Z}^n S^{-1}\Sigma^n_{Z}$ and the displacement $\Sigma_{Z}^n d$ in the phase space. 
We first note that  $U^* = (U^\dagger)^T$ and observe that  $U^* \tau U^{*^\dagger} = (U \tau^T U^\dagger)^T$. The transposition has a simple expression in phase space, namely, for all states $\sigma$,
$W_{\sigma^T}(\xi) = W_{\sigma}(\Sigma^n_{Z}\xi)$ where $\Sigma_{Z}^n = \bigoplus_{k=1}^n \sigma_{Z}$ \cite{Sim00}. 
This leads us to the relation
\begin{align}
	W_{U^*\tau U^{*^\dagger}}(\xi) = W_{\tau}(\Sigma_{Z}^nS^{-1}\Sigma_{Z}^n\xi-\Sigma_{Z}^n d).
\end{align}
To conclude, we observe that $(\Sigma_{Z}^nS^{-1}\Sigma_{Z}^n)^{-1} = \Sigma_{Z}^n S\Sigma_{Z}^n$ is a symplectic matrix since $\Sigma_{Z}^n\Omega(\Sigma_{Z}^n)^T = -\Omega$.

Let us now proceed with the proof of Lemma \ref{lemkey}, which is based on the intrinsic purifications
$\ket{\hat{\psi}_{0}}$  and $\ket{\hat{\psi}_{1}}$ of the $n$-mode states $\rho_{0}$ and $\rho_{1}$. 
We start with the decomposition of $\ket{\psi_b}$ as $\ket{\psi_b}=(U_b^*\otimes U_b)\sum_i\sqrt{p_{b,i}}\ket{i}\ket{i}$. 
Using the basis $\{ U_0\ket{\mathbf{k}}\}_\mathbf{k}$, we can write $\Tr(\sqrt{\rho_{0}}\sqrt{\rho_{1}})$ as
\begin{align}
	\sum_{\mathbf{i,j,k}} \sqrt{p_{0,\mathbf{i}}p_{1,\mathbf{j}}}\;  (\bra{\mathbf{k}}U_{0}^\dagger) U_{0}\proj{\mathbf{i}}U_{0}^\dagger U_{1}\proj{\mathbf{j}} U_{1}^\dagger(U_{0}\ket{\mathbf{k}}),
\end{align}
and the inner product $\braket{\psi_{0}}{\psi_{1}}$ as
\begin{align}
\sum_{\mathbf{i,j}}  \sqrt{p_{0,\mathbf{i}}p_{1,\mathbf{j}}}\; \bra{\mathbf{i}}(U_{0}^\dagger U_{1})^*\ket{\mathbf{j}}\bra{\mathbf{i}}U_{0}^\dagger U_{1}\ket{\mathbf{j}} .
\end{align}
Using $\abs{\braket{\psi_{0}}{\psi_{1}}} = \sqrt{\FF({\psi}_{0},{\psi}_{1})}$ and the definition of $U^*$, a straightforward calculation then shows that
\begin{align}
	\Tr(\sqrt{\rho_{0}}\sqrt{\rho_{1}}) &= \sqrt{\FF({\hat{\psi}}_{0},{\hat{\psi}}_{1}}).
	\label{lemmagic}
\end{align}
Combining Eq.~(\ref{lemmagic}) with inequality~(\ref{battacharyya}) gives
\begin{align}
 1-D(\rho_{0},\rho_{1})\leq\sqrt{\FF({\hat{\psi}}_{0},{\hat{\psi}}_{1})},
\end{align}
which, together with inequality~(\ref{fidelity-distance}), yields
\begin{align}
		D(\hat{\psi}_{0},\hat{\psi}_{1}) \leq \sqrt{2 \, D(\rho_{0},\rho_{1})-D(\rho_{0},\rho_{1})^2}.
\end{align}
This immediately concludes the proof of Lemma~\ref{lemkey}.
\cqfd


\begin{lem}
	\label{lemperfect}
	Let $\ket{\psi_{0}}$ and $\ket{\psi_{1}}$ be $2n$-mode Gaussian states such that $\Tr_{A}\proj{\psi_{0}}=\Tr_{A}\proj{\psi_{1}}$, there exists a Gaussian unitary operator $U$ acting on $n$ modes such that $(U\otimes \I)\ket{\psi_{0}} = \ket{\psi_{1}}$, where $\I$ is the identity on $n$ modes.
\end{lem}
In the discrete-variable case, this is a consequence of the Schmidt decomposition of $\ket{\psi_0}$ and $\ket{\psi_1}$. Here, this role is played by the \emph{normal mode decomposition} \cite{BR03}.
Noting as $\mu_{b}=\left[ \begin{smallmatrix} \mu_b^A \\ \mu_b^B \end{smallmatrix}\right]$ and $\gamma_{b}=\left[ \begin{smallmatrix} \gamma_b^A & C_b \\ C_b^T & \gamma_b^B \end{smallmatrix} \right]$  the first- and second-order moments of $\ket{\psi_{b}}$, the perfectly concealing condition $\Tr_{A}\proj{\psi_{0}} = \Tr_{A}\proj{\psi_{1}}$ implies that $\mu_{0}^{B} = \mu_{1}^{B}$ and $\gamma_{0}^{B}=\gamma_{1}^{B}$. As a consequence, $\gamma_{0}^{A}$ and $\gamma_{1}^{A}$ have the same symplectic spectra, so that, by applying the normal mode decomposition on $\gamma_{0}$ and $\gamma_{1}$ we know that there exist symplectic matrices $S_{b}^{j}$ such that:
\begin{align}
	\gamma_{0} &= (S_{0}^A \oplus S_{0}^{B}) \tilde\gamma  (S_{0}^A \oplus S_{0}^{B})^{t}, \\
	\gamma_{1} &= (S_{1}^A \oplus S_{1}^{B}) \tilde\gamma  (S_{1}^A \oplus S_{1}^{B})^{t}.
\end{align}
$S_{0}^{B}$ and $S_{1}^{B}$ can be chosen equal since $\gamma_{0}^{B}=\gamma_{1}^{B}$. The symplectic matrix $\SS = S_1^A(S_0^A)^{-1} \oplus \I_{2n}$ transforms $\gamma_{0}$ into $\gamma_1$ by acting on Alice's modes only. Similarly, the displacement $\mu_{1}-\SS \mu_{0}$ transforms $\mu_{0}$ into $\mu_{1}$ by acting on Alice's side only, which proves Lemma \ref{lemperfect}. \cqfd

\phantomsection
\addcontentsline{toc}{section}{Perfectly Concealing Protocols}

\emph{Perfectly concealing protocols} --- 
We now turn to the proof of our no-go theorem for Gaussian QBC. 
For perfectly concealing protocols ($\varepsilon$=0), Alice's cheating strategy is well-known: she simply applies an appropriate unitary operation to her half of $\ket{\psi_b}$ between the two stages of the protocol. This allows her to convert $\ket{\psi_0}$ into $\ket{\psi_1}$. In the case of Gaussian QBC, Lemma \ref{lemperfect} implies that this cheating unitary is Gaussian.

\phantomsection
\addcontentsline{toc}{section}{Epsilon-Concealing Protocols}

\emph{$\varepsilon$-concealing protocols } --- 
We now investigate the realistic case where the protocol is not perfectly concealing, which will finally lead us to the proof of Theorem~\ref{thethm}. We want to find an explicit Gaussian $\sqrt{2\varepsilon}$-cheating strategy for Alice against a $\varepsilon$-concealing QBC protocol. 
In the first stage of the protocol, Alice creates the state $\ket{\psi_0}$ and sends $\rho_0$ to Bob. In the second stage, if Alice wants to reveal the bit $0$, she sends her half of $\ket{\psi_0}$ to Bob, while if she decides to reveal the bit $1$, she applies a Gaussian unitary operation to her half of $\ket{\psi_0}$, mapping it to $\ket{\psi^\sharp_1}$, and then sends it to Bob.

As a consequence of Lemma~\ref{lemkey}, there exist Gaussian purifications $\ket{\hat{\psi}_{0}}$ of $\rho_{0}$ and $\ket{\hat{\psi}_{1}}$ of $\rho_{1}$ such that $D(\hat{\psi}_{0}, \hat{\psi}_{1}) \leq \sqrt{2D(\rho_{0},\rho_{1})}$. Moreover $\ket{\hat{\psi}_{0}}$ and $\ket{\psi_{0}}$ (resp. $\ket{\hat{\psi}_{1}}$ and $\ket{\psi_{1}}$) are two Gaussian purifications of the same Gaussian state $\rho_{0}$ (resp. $\rho_1$), so that, according to Lemma~\ref{lemperfect}, there exists a Gaussian unitary operator $U_0$ (resp. $U_1$) such that $(U_0\otimes\I)\ket{\psi_0}=\ket{\hat{\psi}_0}$ (resp. $(U_1\otimes\I)\ket{\psi_1}=\ket{\hat{\psi}_1}$). We note $\ket{\psi^\sharp_1} = (U_{1}^{-1}U_0\otimes\I)\ket{\psi_0} = (U_{1}^{-1}\otimes\I)\ket{\hat{\psi}_0}$. By unitary invariance of the trace distance, one has $D(\psi^\sharp_1,\psi_{1})=D(\hat{\psi}_0,\hat{\psi}_{1})$. Thus, for $\varepsilon$-concealing protocols, we have $D(\psi^\sharp_1,\psi_{1})\leq\sqrt{2\varepsilon}$, which concludes the proof of Theorem~\ref{thethm}.\cqfd

We have thus obtained a stronger result than the standard no-go theorem since we have shown that QBC remains impossible even if Alice and Bob are restricted to manipulate Gaussian states. Although Lemma~\ref{lemkey} can be seen as a weak version of Uhlmann's theorem in the sense that the intrinsic purification does not reach Uhlmann's bound, it is sufficient here because the quantities of interest in terms of guessing probability are not changed. Interestingly, the question of whether the purifications that saturate Uhlmann's bound could both be chosen Gaussian if the states are Gaussian is still open  (although partial results in this direction have been obtained in \cite{MM07}). 
Note also that we have an explicit construction of Alice's cheating purifications for any CV QBC protocol, Gaussian or not. This is done by noting that the Gaussian constraint can be relaxed in the proof of Lemma~\ref{lemkey}, and that Lemma~\ref{lemperfect} can be replaced by the usual Schmidt decomposition. 


\phantomsection
\addcontentsline{toc}{section}{CBH Theorem}

\emph{CBH theorem} --- 
Consider the subset of quantum mechanics where only Gaussian states and operations are allowed. As a result
of our no-go theorem, this Gaussian model forbids bit commitment while it allows unconditional secret key distribution \cite{RC09}.
Interestingly, however, it is stricly included in quantum mechanics since, for instance, Bell inequalities cannot be violated with Gaussian states and measurements. This contradicts the Brassard-Fuchs conjecture. Furthermore, 
according to the CBH theorem \cite{CBH03}, quantum mechanics can be rederived from the sole assumptions that signalling, broadcasting, and bit commitment are impossible in Nature. While this idea is very appealing, the Gaussian model again provides a natural counter-example to it.
The reason is that the CBH theorem actually requires the further assumption that the physical description of Nature is done within the framework of $C^*$-algebras (Spekkens had found a toy model compatible with CBH but distinct from quantum mechanics \cite{Spe07}, but
ours is physically better grounded).  



\phantomsection
\addcontentsline{toc}{section}{Conclusion}

\emph{Conclusion} --- 
We have addressed continuous-variable quantum bit commitment, and have proven a strong version of the standard no-go theorem in which Alice and Bob are restricted to Gaussian states and operations. Our proof is based on a Gaussian purification of Gaussian states, eliminating the need for Uhlmann's theorem. Note that Bob is not restricted to Gaussian measurements at the last stage of the protocol, which may make him more powerful than in a fully Gaussian protocol. Even then, Alice can always perform a Gaussian cheating strategy. This leaves open, however, the possible existence of non-Gaussian QBC protocols that could be secure against Gaussian attacks. More fundamentally, we have exhibited a physically motivated counter-example to the attempts at rederiving quantum mechanics from first principles.


\begin{acknowledgments}
We thank R. Garc\'ia-Patr\'on, A. Grinbaum, and S. Pirandola  for fruitful discussions. We acknowledge financial support of the European Union under the FET projects COMPAS (212008) and QAP (015848), of Agence Nationale de la Recherche under projects SEQURE
(ANR-07-SESU-011-01), QRAC (ANR-08-EMER-012), CRYQ (ANR-09-JCJC-560290), and of Brussels-Capital Region under project CRYPTASC.
\end{acknowledgments}

\bibliographystyle{unsrt}

\end{document}